\documentclass[aps,pre,showpacs,twocolumn]{revtex4}
\usepackage{graphicx,verbatim}

\newcommand{\req}[1]{(\ref{#1})}
\newcommand{\be}{\begin{equation}}
\newcommand{\ee}{\end{equation}}
\newcommand{\bea}{\begin{eqnarray}}
\newcommand{\eea}{\end{eqnarray}}

\newcommand{\avg}[1]{\langle{#1}\rangle}

\newcommand{\BE}{\begin{eqnarray}}
\newcommand{\EE}{\end{eqnarray}}
\newcommand{\BEn}{\begin{eqnarray*}}
\newcommand{\EEn}{\end{eqnarray*}}
\newcommand{\barr}{\begin{array}}
\newcommand{\earr}{\end{array}}

\newcommand{\bit}{\begin{itemize}}
\newcommand{\eit}{\end{itemize}}
\newcommand{\bc}{\begin{center}}
\newcommand{\ec}{\end{center}}
\newcommand{\ben}{\begin{enumerate}}
\newcommand{\een}{\end{enumerate}}

\begin{document}

\title{Structure-preserving desynchronization of minority games}
\author{Giancarlo Mosetti}
\affiliation{D\'epartement de physique, Universit\'e de Fribourg, P\'erolles, 1700 Fribourg, Switzerland}
\affiliation{ISI Foundation, Vialle Settimio Severo 65, 14011 Turin, Italy}
\author{Sorin Solomon}
\affiliation{The Racah Institute of Physics, Hebrew University,, Jerusalem,  91905, Israel}
\affiliation{ISI Foundation, Vialle Settimio Severo 65, 14011 Turin, Italy}
\author{Damien Challet}
\affiliation{D\'epartement de physique, Universit\'e de Fribourg, P\'erolles, 1700 Fribourg, Switzerland}
\affiliation{ISI Foundation, Vialle Settimio Severo 65, 14011 Turin, Italy}
\email{damien.challet@unifr.ch}
\date{\today}


\begin{abstract}
Perfect synchronicity in $N$-player games is a useful theoretical dream, but communication delays are inevitable and may result in asynchronous interactions. Some systems such as financial markets are asynchronous by design, and yet most theoretical models assume perfectly synchronized actions. We propose a general method to transform standard models of adaptive agents into asynchronous systems while preserving their global structure under some conditions. Using the Minority Game as an example, we find that the phase and fluctuations structure of the standard game subsists even in maximally asynchronous deterministic case, but that it disappears if too much stochasticity is added to the temporal structure of interaction. Allowing for heterogeneous communication speeds and activity patterns gives rise to a new information ecology that we study in details.
\end{abstract}

\maketitle

\section{Introduction}

When a large number of agents taking part in a multi-player game submit their actions to a central authority (game server, financial market), the times at which their actions become effective are likely to differ  because of reaction times, transmission delays or backlogs at the central server. If the resulting delay is sufficiently large, the synchronicity of actions and payoffs is not a reasonable assumption anymore. Yet the immense majority of the literature on games and agents assumes perfect synchronicity (see however e.g. \cite{HubermanAsyncPNAS,LagunoffAsync,VegaAsync} for some notable exceptions).

This is problematic in the modelling of many systems, among which the numerous financial markets where the actions of agents are discrete in time and asynchronous. Most financial market models aggregate traders' actions over a given period in one time step. Unless it corresponds to sensible time periods, such as one trading day, this approach is rather artificial. While time coarsening simplifies the description of market dynamics, speculation cannot be modelled by including all the agents' actions in one time step: one does not make money with a single transaction, i.e. in a single time step. In addition, the emergence of large price and volume fluctuations must also be explained in an asynchronous setting. Finally asynchronicity also originates from the heterogeneity of time scales of market participants, which is fat-tailed, possibly a power-law \cite{ZumbachLynch,LilloUtility,CMor08}.

This raises fundamental issues regarding macroscopic synchronization, especially in the case of coordination and cooperation.  Here we modify the well-understood minority game \cite{CZ97} by introducing a tunable time delay between the submission of a bid and its actual influence on the global outcome, and a tunable playing frequency.

Remarkably the structure of mean-field models such as the MG is preserved; as a consequence there is hope that the resulting asynchronous interaction somehow still belongs to the mean-field category, hence, that the powerful methods from statistical mechanics that solve the original model \cite{MGbook,CoolenBook} can be generalized.

\section{Desynchronizing global games}

Let us consider a global synchronous game. Each agent $i=1,\cdots,N$ takes action $a_i(t)$ at time $t$. His payoff is a function of his own action and of the global action of all the agents $A(t)=\sum_{j=1}^N a_j(t)$: everything can be written as a function of time $t$ only.

\begin{figure}
\centerline{\includegraphics[width=0.4\textwidth]{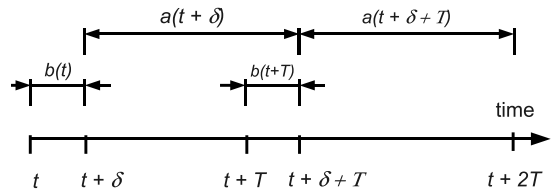}}
\caption{Schematic explanation of playing period $T$, time delay $\delta$, bids $b(t)$ and actions $a(t+\delta)$.}
\label{fig:schema}
\end{figure}

A way to desynchronize the agents while preserving the global structure of the game is to think in terms of delays: if agent $i$ submits his action at time $t$, the latter becomes effective (i.e. is incorporated into $A$) only at time $t+\delta$. But at time $t+\delta$, $A(t+\delta)$ only contains the actions effective at that time, that is, does not contain the action of some other agents that are still being transmitted or thought about. Thus, the payoffs of the agents at this time do not reflect the most recent actions, but only the last known actions. Differentiating between actions sent but not effective yet, thereafter called bids and denoted by $b_i(t)$, and actual actions $a_i(t)$ makes it easy to introduce asynchronous actions while keeping intact the payoff structure, as will be shown in the next section (see Fig.\ \ref{fig:schema}).

Mathematically, agent $i$ submits his bid $b_i(t)$; between times $t$ and $t+\delta-1$, his last known action is unchanged; at time $t+\delta$, the last bid becomes effective, i.e. $a_i(t+\delta)=b_i(t)$ and of course $A(t+\delta)=\sum_i a_i(t+\delta)$; agent $i$ then receives his payoff that depends on $a_i(t+\delta)$ and $A(t+\delta)$. In this way, a non-trivial structure of overlapping bids and actions can be built.

The temporal structure of the game is completely specified by assuming that agent $i$ is active at times $t_i^{(n)}$, $n>0$. For instance, $t_i^{(n)}nT_i+\phi_i$ where $n$ is an integer. The maximally asynchronous case corresponds to $T_i=N$ and $\phi_i=i$: only one agent is active at each time step. Alternatively, an agent may be active with some probability at each time step, thereby removing the rigid structure imposed by periodic $t_i^{(n)}$.

\section{Example: Minority Games}

The minority game (MG thereafter) is a prototype model of global competition between adaptive agents \cite{CZ97,MGbook}. Well-understood \cite{MGbook,CoolenBook}, it provides an ideal test-bed for new ideas and extensions. At the same time, it has highly non-trivial and characteristic fluctuations structure and  phase transition. The existence of a phase transition with symmetry breaking \cite{Savit,CM99} is robust with respect to a surprising number and types of modifications \cite{MGbook,dMG08}; as such, if present in modified games, it is a signature that the original dynamics has not been overly altered.

The aim of the agents is, as the name of the game implies, to be in the minority: in the original game, agent $i$ takes action $a_i(t)\in\{-1,1\}$; $A(t)$ is defined as above and is positive if the majority chose $+1$ and vice-versa. The payoff of agent $i$ is $-a_i(t)A(t)$: those who happen to be in the minority are rewarded.

The various types of minority games found in the literature differ mostly in their learning and decision mechanisms. The original one is defined as follows: agents are fed with the last $m$ winning decisions, a bitstring called history and denoted by $\mu(t)$. Each agent has a set of $S$ strategies, i.e., of predefined ways to react to all possible public pieces of information. Denoting the strategies of agent $i$ by $a_{i,s}$, $s=1,\cdots,S$, one can rewrite $A(t)=\sum_i a_{i,s_i(t)}^{\mu(t)}$ where $s_i(t)$ is the strategy trusted by agent $i$ at time $t$.

Which strategy to choose is determined by reinforcement learning. To this effect, since the agents gather information about the world through the use of their strategies, they store experience about the past in virtual performance scores of their strategies that evolve according to
\be\label{eq:payoff}
U_{i,s}(t+1)=U_{i,s}(t)-a_{i,s}^{\mu(t)}A(t)
\ee
The agents choose their best strategy at time $t$. In other words, $s_i(t)=\arg\max_s U_{i,s}(t)$.

When desynchronizing the game, a slight complication with respect to game histories arises: whereas in perfectly synchronized games $\mu(t)$ is the same for all the agents, this cannot hold anymore in asynchronous settings since all the agents do not see the same $A$ when they receive a payoff. Thus each agent has his own history of the game, which encodes the past $m$ right choices for him.

By construction $\avg{A}=0$. We shall be interested in the fluctuations of the global outcome $\sigma^2=\avg{A^2}$ which quantify the degree of coordination of the players, the benchmark being the random outcome $\sigma^2=N$. Predictability must be measured at the individual level: one defines the conditional average of the attendance from the point of view of agent $i$, that is, conditional to his histories, which we denote as  $\avg{A^{(i)}|\mu_i}$, and the predictability as seen by this agent $H_i=\sum_{\mu_i}\avg{A^{(i)}|\mu_i}^2/P$, where $P=2^M$, while the average predictability is $H=\sum_iH_i/N$.

Predictability $H>0$ corresponds to a (conditional) symmetry breaking between the two choices. The standard game is characterized by a predictable phase ($H>0$) for $\alpha=P/N>\alpha_c$, an unpredictable phase ($H=0$, $\alpha<\alpha_c$) and critical point $\alpha_c=0.3374\dots$ where $H\to0$ and $\sigma^2/N$ reaches a minimum \cite{Savit,CM99,MGbook,CoolenBook}. The presence of this phase transition is acknowledged to be robust with respect to many modifications of the game, except when all the agents take into account their impact on the game by removing their contribution to $A$, i.e., by replacing $A$ with $A_{-i}=A-a_i(t)$ in Eq \req{eq:payoff} \cite{CMZe00,dMG08}. However, little is known about the importance of the synchronization of histories with respect to the existence of the phase transition. Local games are synchronized MGs where an agent plays against his neighbors, giving rise to partially spatially overlapping games, hence, histories \cite{Kalinoswki,DeLosRiosMG}, whereas the overlaps are in time in our case.

Another issue is the information ecology: given the fact that predictability is easily measured,  MGs give insights on who exploits whom \cite{MMM}, that is, in this case, on the risk associated with delays, for instance. Since the MG is a negative sum game, making an average positive gain is hard; it is  only achieved in the original game by some agents that can exploit very efficiently the predictability left by other agents, such as e.g. some of those able to settle on one strategy in the standard game (frozen agents) \cite{Savit,CM99,MMM}, speculators feeding on producers, insider trading or a longer history length deep in the unpredictable phase \cite{MMM}.  Recently two of us designed and studied synchronous MGs where the agents had heterogeneous time scales \cite{MoCZ05}.  Here we shall characterize the importance of communication delays and frequency of play.

\subsection{Results}

Numerical simulations are about $N$ times slower than those of the usual MGs, as one effective time-step from a measurement point of view ends when all the agents have updated their actions. Even worse, the interesting regions are found for quite low $\alpha=P/N$, which makes computations even slower. This unfortunately limits the system sizes one can study with current computers to $M=4$, i.e. $P=16$. Indeed one run for the maximally asynchronous case at $P=32$ and $P/N=0.01$ needs 150 minutes a modern computer (Core 2 duo, 2GHz) , hence averaging over 200 samples requires 20
days just for this point. Fortunately, $M=4$ yields good enough results, as discussed in the concluding section.

\subsubsection{Maximally asynchronous game}
\label{sec:maxasync}
Assuming that $t_i^{(n)}=nT_i+\phi_i$ as above and setting $T_i=T=N$ and $\phi_i=i$, there is only one active agent at each time step, hence $A$ changes at most by $2$ in a time step. In this case, the histories $\mu_i$ of agents 1 and  $N/2$ will likely differ, unless the dynamics of $A$ has a memory longer that  $N/2$.

\begin{figure}
\centerline{\includegraphics*[width=0.4\textwidth]{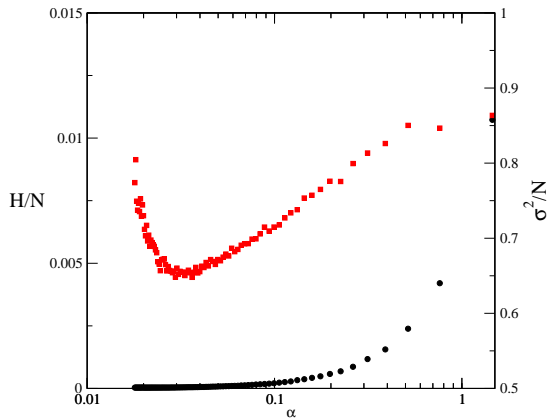}}
\caption{Fluctuations $\sigma^2/N$ (red squares; right scale) and predictability $H/N$ (black circles; left scale) of the maximally asynchronous MG as a function of $\alpha=P/N$. $M=4$ ($P=16$), $S=2$, $400NP$ iterations,  $200NP$ time steps until equilibration, averages over 200 samples.}
\label{fig:maxasyncMGs2H}
\end{figure}

Plotting as usual $\sigma^2/N$ and $H/N$ reveals that this system undergoes the same phase transition as the original minority game (see Fig.\ \ref{fig:maxasyncMGs2H}). This means that the desynchronization we propose has the remarkable property to keep the global structure of the game unchanged, while allowing for extreme asynchronicity. It is notable that the predictability is orders of magnitude smaller than $\sigma^2/N$, which is to be expected since the very long time delays $T_i$ of the agents dilutes information; the order of magnitude of the fluctuations on the other hand is unchanged, since $A\propto \sqrt{N}$.

The auto-correlation of $A$ reveals a complex pattern, similar to that seen in Fig.\ \ref{fig:corrHetTimeDelays}: since $A$ is only possibly changed by only one agent at a time, it displays persistence for $O(N/2)$ time steps. After this decay, it shows on average negative auto-correlation, as can be expected in a minority game where every deviation from $A=0$ tends to be cancelled by adaptive agents. The additional oscillations are of period $N$.

\subsubsection{Heterogeneous time delays}

The pattern of auto-correlation of $A$ suggest that playing with a high frequency allows one to take profit from the persistence of $A$.

\begin{figure}
\centerline{\includegraphics*[width=0.4\textwidth]{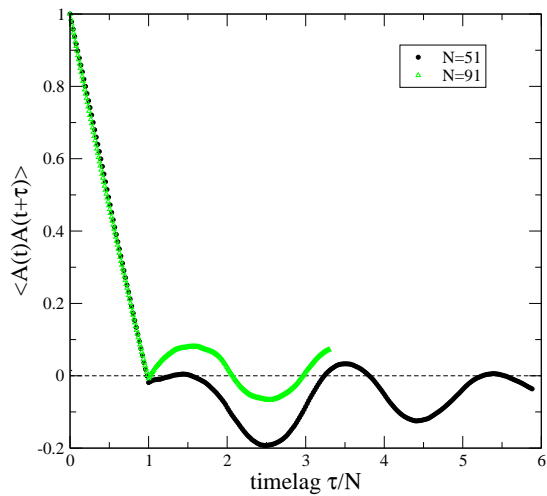}}
\caption{Autocorrelation of $A(t)$ in a population with heterogeneous time delays $T_i$;  $M=4$ ($P=16$), $N=51$, $S=2$, $400NP$ iterations,  $200NP$ time steps until equilibration.}
\label{fig:corrHetTimeDelays}
\end{figure}

\begin{figure}
\centerline{\includegraphics*[width=0.4\textwidth]{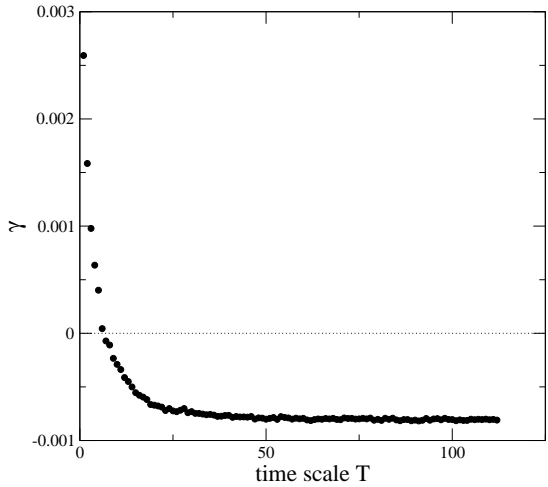}}
\caption{Average gain of agents with a given time delay $T_i$ versus $T_i$  $M=4$ ($P=16$), $N=101$, $S=2$, $400NP$ iterations,  $200NP$ time steps until equilibration, averages over $10^5$ samples.}
\label{fig:gainsHetTimeDelays}
\end{figure}

Drawing at random $T_i$ from $\{1,\cdots,\rho N\}$ and setting $\delta_i=T_i$, allows one to study the respective gains associated with a time scale denoted by $\avg{g|T}=\avg{g|\delta}$, in order words, the information ecology that arises from being active more often and having a shorted delay in an asynchronous setting. Figure \ref{fig:gainsHetTimeDelays} shows that in asynchronous situations being faster is an clear advantage, a few of the agents reaping even positive gains.

Positive gains come quite peculiarly for minority games from the unconditional persistence of $A$. Computing the autocorrelation.$\avg{A(t)A(t+\tau)}$ (Fig.\ \ref{fig:corrHetTimeDelays}) shows a memory of $A$ lasting $N/2$ time steps, which corresponds exactly to the point where the losses of the agents saturates in Fig \ref{fig:gainsHetTimeDelays}. Earning a positive gain, though, requires enough persistence to overcome the cost of play the MG, a negative sum game.

\subsubsection{Activity frequency and time delays}

\begin{figure}
\centerline{\includegraphics*[width=0.4\textwidth]{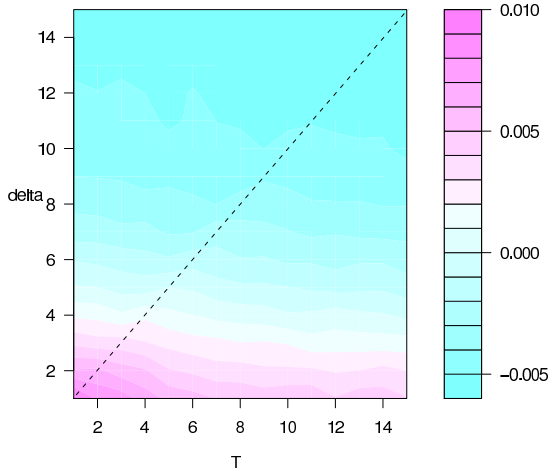}}
\caption{Contour plot  $M=4$ ($P=16$), $N=101$, $S=2$, $400NP$ iterations,  $200NP$ time steps until equilibration, averages over $10^5$ samples.}
\label{fig:contour_plot}
\end{figure}

In the two previous subsections, one assumed that the frequency of activity of a given agent is exactly equal to his time delay. This restriction is unrealistic and can be lifted easily: assume that an agent plays every $T_i$ time steps and has a time delay of $\delta_i$. The delay $\delta_i$ may happen to be larger than $T_i$: if an agent is unfortunate enough to have a larger delay than activity frequency, he waits for $\delta_i$ time steps to receive his payoff and sends a new bid at the next possible activity time $t_i^{(n)}$.

The average gains conditional on $\delta$ and $T$, denoted by $\avg{g|\delta,T}$,  reveal in more details the information ecology of asynchronicity: Figure \ref{fig:contour_plot} reports a contour plot of $\avg{g|\delta,T}$; $T$ and $\delta$ play a similar role: the more frequently one is active, the more one can profit from the persistence of $A$, i.e., from the slowness of other players, as before. But $\delta$, which can be seen as the quickness of reaction to new information,  is expectedly the most relevant parameter: it mainly controls whether one obtains a positive or negative gain, except around $\delta=4$, where $T$ plays this role; the gain decreases as $T$ increases, which is consistent with the results of the previous subsection. The last feature of the plot is the slight bump at $\delta=T$, which is due to the fact that the players who have $\delta=T+1$ have effectively $T'=2T$, hence the slight decrease of average gain.

\subsubsection{Poissonian activity}

\begin{figure}
\centerline{\includegraphics*[width=0.4\textwidth]{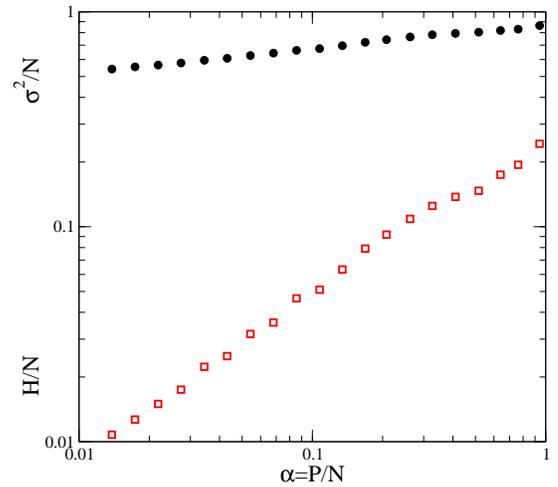}}
\caption{Predictability $H/N$ and fluctuations $\sigma^2/N$ as a function of $\alpha=P/N$ for Poissonian activity ($\theta=1$). $M=4$ ($P=16$), $S=2$, $400NP$ iterations,  $200NP$ time steps until equilibration, averages over 200 samples.}
\label{fig:s2H_prob}
\end{figure}

\begin{figure}
\centerline{\includegraphics*[width=0.4\textwidth]{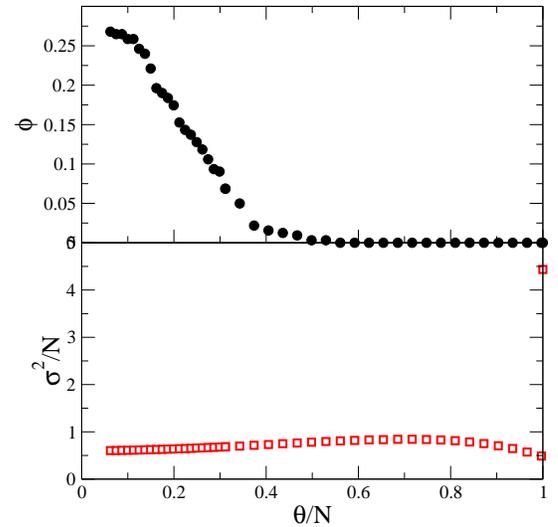}}
\caption{Fraction of frozen agents $\phi$ and fluctuations $\sigma^2/N$  as a function of $\theta/N$ for Poissonian activity for a single realisation of the disorder. $M=4$ ($P=16$), $S=2$, $80000NP/\theta$ iterations,  $4000NP/\theta$ time steps until equilibration.}
\label{fig:phi_theta}
\end{figure}

Finally, let us relax completely the periodicity of agent activity: each agent $i$ plays now at time $t$ with probability $\theta/N$; the standard MG is recovered when $\theta=N$. We first take $\theta=1$, which is stochastically equivalent to the setup of section \ref{sec:maxasync}, except that the temporal interaction has no more structure. Figure \ref{fig:s2H_prob} suggests that the phase transition disappears completely, at least  for the range of parameters we swept over (we tried also tried individual runs at $\alpha=0.001$ that confirm this result).

This means that at least some structure of overlap between the histories of the agents is a crucial ingredient of the phase transition. If the phase transition is resilient to some stochastic perturbation of the personal history update temporal structure, one should find a critical $\theta_c$ for which a system placed in the symmetric phase undergoes a phase transition when sweeping over $\theta$. The best way to test for a phase transition is to plot the fraction of frozen agents, $\phi$, defined as the fraction of agents which played the same strategy during the last half of the time steps \cite{CM99}. The only situation for which $\phi=0$ can occur is when $H=0$.  Figure \ref{fig:phi_theta} shows that the system stays in the symmetric phase as long as $\theta\ge\theta_c/N\simeq 0.5$ and then $\phi$ increases rapidly. The fluctuations drop discontinuously at $\theta=1$: anergordicity is broken by stochasticity, but the system stays in a symmetric phase.

\section{Discussion}

We have proposed a powerful and generic way to desynchronize global games while preserving their structure. The resulting asynchronicity allows for the study of the effect of time delays and playing frequency in principle in any $N$-player game.

Desynchronized MGs provide yet another example of the robustness of the phase transition in MGs (see the many other phase transition-preserving modifications listed in \cite{MGbook}). We did not study in detail the location of this phase transition;  additional lengthy numerical simulations are needed to study this point in detail, in particular as a function of $T$.

 The fact that the phase transition disappears in the presence of strong enough stochastic desynchronization is a clue that this robustness relies on the temporal structure of individual history updates, as also confirmed by the fact that history-less games ($P=1$) \cite{MC00,CAM08} do not reproduce the delay information ecology found in the present study. Interestingly, recent work on the Prisoner's Dilemma showed that similar Poissonian desynchronization leads to a first order phase transition at finite activity frequency \cite{SaifGadePD}.

Although we had to consider quite small systems ($M=4$), the results presented here will not change qualitative when simulating much larger systems is doable. Finite size effects of minority games are well-studied \cite{MGbook};  two points are crucial: i) all the macroscopic variables depend only on $\alpha=2^M/N$, up to finite size effects; ii) the existence of the phase transition is found for all $M>1$. It should be noted that macroscopic variables ($H/N$, $\sigma^2/N$) of the MG with real histories may have peculiar finite size effects for very small $M$ because of the De Bruijn graph on which bitstrings of length $M$ live; when $M\ge4$, its complexity is large enough to prevent the histories of the game to be stuck in a small trivial subgraph.

The familiar phase structure of asynchronous games suggests to solve this kind of games with mathematical methods that have been successfuly applied to the original MGs and some of its extensions \cite{MGbook,CoolenBook}. Two new problems arise: first actions are delayed, which makes the whole calculus more complex; the second problem comes from the fact that each agent has his own real history: solving the standard MG with real (global) histories was a mathematical {\em tour de force} \cite{CoolenRealHist,CoolenBook}; solving systems with individual histories will need an even more impressive feat. Nevertheless, there is nothing {\em in principle} that makes the computation infeasible, although it will be much more complex. We hope that such future work will reveal the inner dynamical effects of asynchronicity.

\bibliographystyle{unsrt}
\bibliography{biblio}

\end{document}